\documentclass[twocolumn,aps,prl,floatfix]{revtex4-1}
\usepackage{graphicx}
\usepackage{amsmath,amssymb}
\usepackage{subfigure,overpic}
\usepackage{hyperref}
\usepackage[usenames]{color}
\usepackage{textcomp}
\usepackage{color}

\begin{document}

\title {Tunable mechanical and thermal properties of ZnS/CdS core/shell nanowires}

\author {Taraknath Mandal, Chandan Dasgupta and Prabal K. Maiti}

\affiliation{Center for Condensed Matter theory, Department of Physics, Indian Institute of Science, Bangalore 560012, India}

\begin{abstract}
Using all atom molecular dynamics (MD) simulations, we have studied the mechanical properties of ZnS/CdS core/shell nanowires. Our results show that the coating of a few atomic layer CdS shell on the ZnS nanowire leads to a significant change in the stiffness of the core/shell nanowires compared to the stiffness of pure ZnS nanowires. The binding energy between the core and shell region decreases due to the lattice mismatch at the core-shell interface. This reduction in binding energy plays an important role in determining the stiffness of a core/shell nanowire. We have also investigated the effects of the shell on the thermal conductivity and melting behavior of the nanowires. 
\end{abstract} 

\maketitle

\section {Introduction}
One dimensional nanowires have attracted considerable amount of research interest because of their potential application in the field of modern nanotechnology. The most important characteristic of these nanostructures is very high surface to volume ratio. As a result, surface effects which are negligible in the bulk material, can lead to a significant changes in the mechanical~\cite{agrawal2008elasticity, chen2006size}, thermal~\cite{martin2009impact, kulkarni2006size}, electronic~\cite{banerjee2000effect, baskoutas2006size, li2001band} and structural~\cite{kondo1999thickness, kondo1997gold, hall1991multiply, hasmy2002thickness, mandal2012strain} properties of the nanostructures which are very different from their bulk counterparts. The properties of the nanoparticles can be tuned by functionalizing the nanoparticle by other organic/inorganic materials. Core/shell nanowires are one kind of functionalized nanowires where the surface region of a pure nanowire is coated by a thin shell of different materials. In recent years, many experimental investigations have been carried out to synthesize and characterize different kinds of core/shell structures such as ZnO/ZnS~\cite{lu2009zno}, ZnO/CdS, ~\cite{gao2005sonochemical} GaN/GaP, ZnO/TiO2~\cite{law2006zno, greene2007zno}, CdSe/CdS~\cite{pan2005synthesis, li2003large}, PbSe/CdSe~\cite{zhang2010formation, zhang2011pbse}, ZnS/CdS~\cite{liu2011low, qu2012zns, pouretedal2011photodegradation, jung2007synthesis, murugadoss2012structural}, CdS/ZnS~\cite{liu2011low, wang2009synthesis, manna2002epitaxial}, Ge/Si ~\cite{wingert2011thermal}. Both the classical molecular dynamics (MD) and density functional theory (DFT) methods have been used extensively to study the electronic~\cite{huang2007density, peng2010electronic, saha2013tuning, sadowski2010core, musin2005structural, pekoz2009bare, pekoz2011band}, optical~\cite{huang2007density, schrier2007optical} and thermal ~\cite{hu2010significant, chen2012impacts, markussen2012surface, hu2011thermal, huang2014tunable, zeng2014high, huang2012pt, huang2013insight, huang2013thermal, huang2012two, cheng2013molecular, song2010molecular} properties of the core/shell nanostructures. In contrast, mechanical properties of the core/shell nanostructures have so far not been investigated in details. Mechanical properties of nanostructures have a strong influence on their electronic ~\cite{balamurugan2005evidence, tekleab2001strain, smith2008tuning} and optical~\cite{smith2008tuning, smith1987optical} properties. Hence, knowledge of the mechanical and thermal properties of the core/shell nanostructures is very important for building efficient nanodevices. \\

In this paper, we have used fully atomistic classical MD simulations to study the mechanical properties of ZnS/CdS core/shell nanowires. ZnS and CdS nanowires are of particular interest because these wide band gap semiconducting materials are promising candidates for various applications, such as, solar cell~\cite{oladeji2005synthesis, lee2003growth, burton1976znxcd1}, sensors~\cite{koneswaran2009cysteine, pardo2003acetylcholine}, photodetectors~\cite{hayden2006nanoscale}, and electronic devices. We show that the Young's modulus of wurtzite ZnS nanowires can be tuned significantly by introducing a thin layer of CdS shell on the surface of the ZnS nanowire. We have also used first principle DFT method to verify the accuracy of some of the results obtained from the classical MD simulations. To check the influence of this shell on the thermal properties, we have investigated the thermal conductivity and melting behavior of these core/shell nanowires. We find that both the thermal conductivity and melting temperature of the ZnS/CdS core/shell nanowires are significantly lower than those of the pure ZnS nanowires. 

\section {Methodology}
All nanowires used in this study are grown along the [0001] direction. Initial configurations of the nanowires are generated using Cerius2 ~\cite{cerius2}. The core regions of the core/shell nanowires are cut from a super cell of the wurtzite ZnS crystal. After constructing the core region, the CdS shell of desired width was built on top of the core region. We denote the core/shell nanowire as shell1, shell2, shell3, shell4 and shell5 depending on the number of CdS layer on top of the ZnS core. For example, shell1 nanowire has one CdS layer on top of the ZnS core and shell2 has two layers of CdS on top of ZnS core and so on. The total number of layers is kept fixed for each nanowire. The equilibrated structure of a core/shell nanowire is shown in figure 1. Interatomic interactions between Zn-Zn, Zn-S, Zn-Cd, Cd-S, Cd-Cd and S-S are described by Lennard Jones (LJ) and Coulomb potential. Zn, S and Cd atoms have charges of +1.18e, -1.18e and +1.18e, respectively. The potential parameters for ZnS and CdS and the partial charges of the atoms are taken from the reference~\cite{grunwald2012transferable}. These potential parameters were shown to describe the lattice and elastic constants and phonon dispersion of ZnS and CdS accurately. Buckingham potential ~\cite{wright2004interatomic} also has been used in some of the studies to describe the short range van der Waals interactions of the ZnS and CdS systems. However, this potential has not been used so far to study multi- component systems (hetero-structures). We have used this particular LJ potential since these ZnS and CdS parameters are specifically designed to be compatible with each other so that they are able to simulate mixture of ZnS and CdS ~\cite{grunwald2012transferable}. These parameters are also used to study different properties of CdSe/ZnS, CdSe/CdS hetero-structures~\cite{grünwald2012metastability, eshet2013electronic}. Periodic boundary condition is enforced along all three directions during the simulation. The lengths of the simulation box along the growth direction (Z direction) of the nanowire is the same as that of the nanowire ($\sim 12.3 nm$) but lengths along the other two directions (X and Y directions) are 15 nm, considerably larger than the diameters of the nanowires so that the interactions between the nanowire and its periodic images are negligible. A similar protocol was used in our previous~\cite{mandal2012mechanical, mandal2012strain} studies where we have investigated the mechanical properties of one dimensional structures. Particle-Particle-Mesh-Ewald method ~\cite {hockney1988computer} is used to compute the long range electrostatic interactions. The initially built structures were first equilibrated at constant pressure for 500 ps and then at constant volume for another 500 ps at 300 K temperature before subjecting them to tensile load. The Debye temperature of the ZnS and CdS are $277\pm7$ K~\cite{blattner1972x} and $210\pm20$ K~\cite{look1972nuclear}, respectively. So we expect all the vibrational modes to be excited while doing MD simulation at 300 K. The tensile process was carried out by separating a few of the top and bottom layers of atoms, considering them to be two rigid blocks, in a stepwise manner. For each step, the nanowires were stretched by 0.3 $\AA$ and equilibrated for 100 ps.

\begin{figure}
\centering
       \centering
        \subfigure
        {
        \includegraphics[height=45mm, width=80mm]{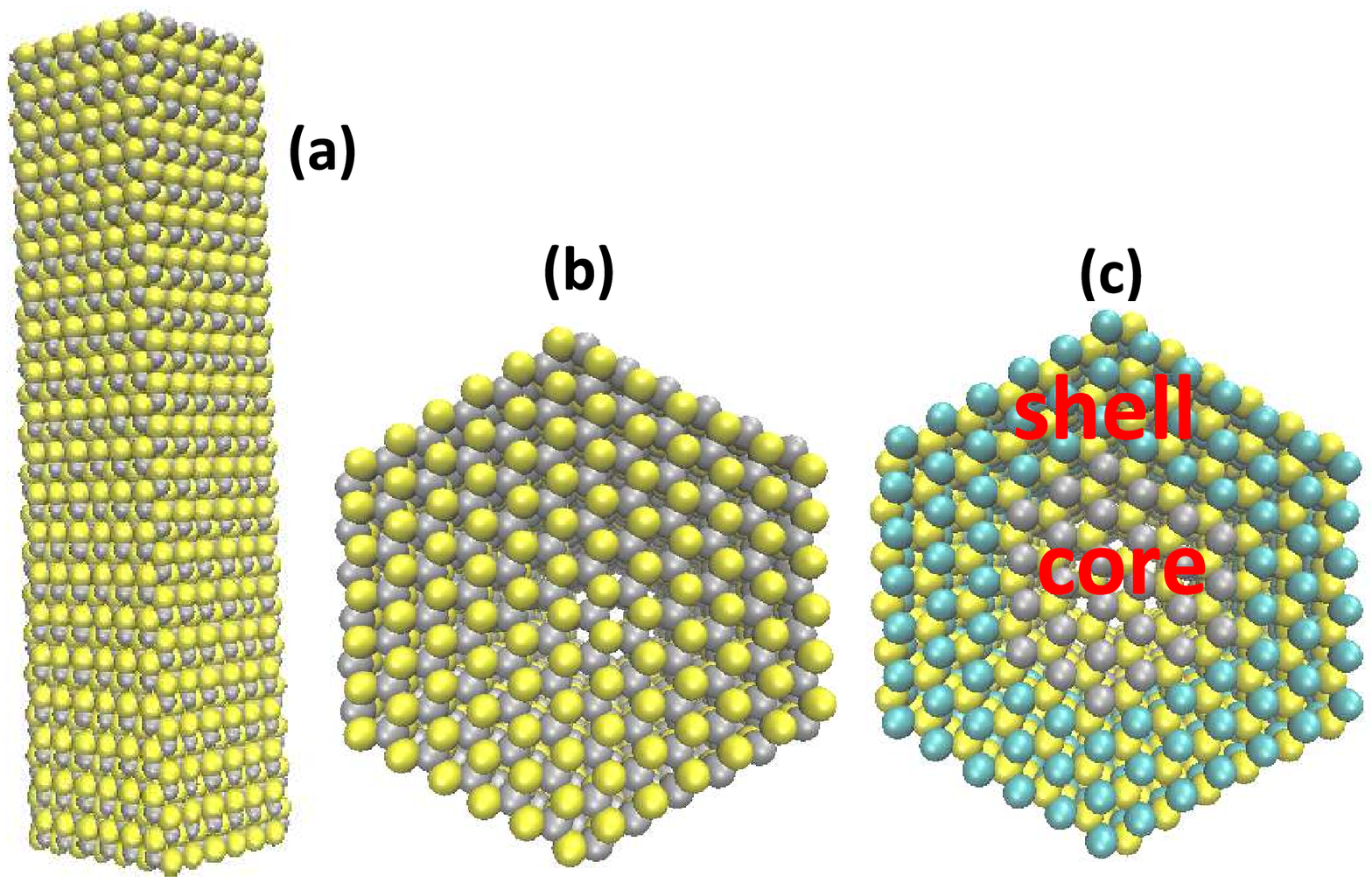}
        } 
\caption{(Color online) (a) Equilibrated structure of a ZnS nanowire. (b) and (c) are cross-sectional views of a ZnS and a shell3 ZnS/CdS nanowire, respectively. Gray, Yellow and blue colors represent Zinc, sulphur and Cadmium atoms, respectively. We call a core/shell nanowire as shell3 when the three outermost layers of pure ZnS nanowire are replaced by CdS atoms.}
\end{figure}

\begin{figure}
\centering
        \subfigure
        {
        \includegraphics[height=40mm, width=80mm]{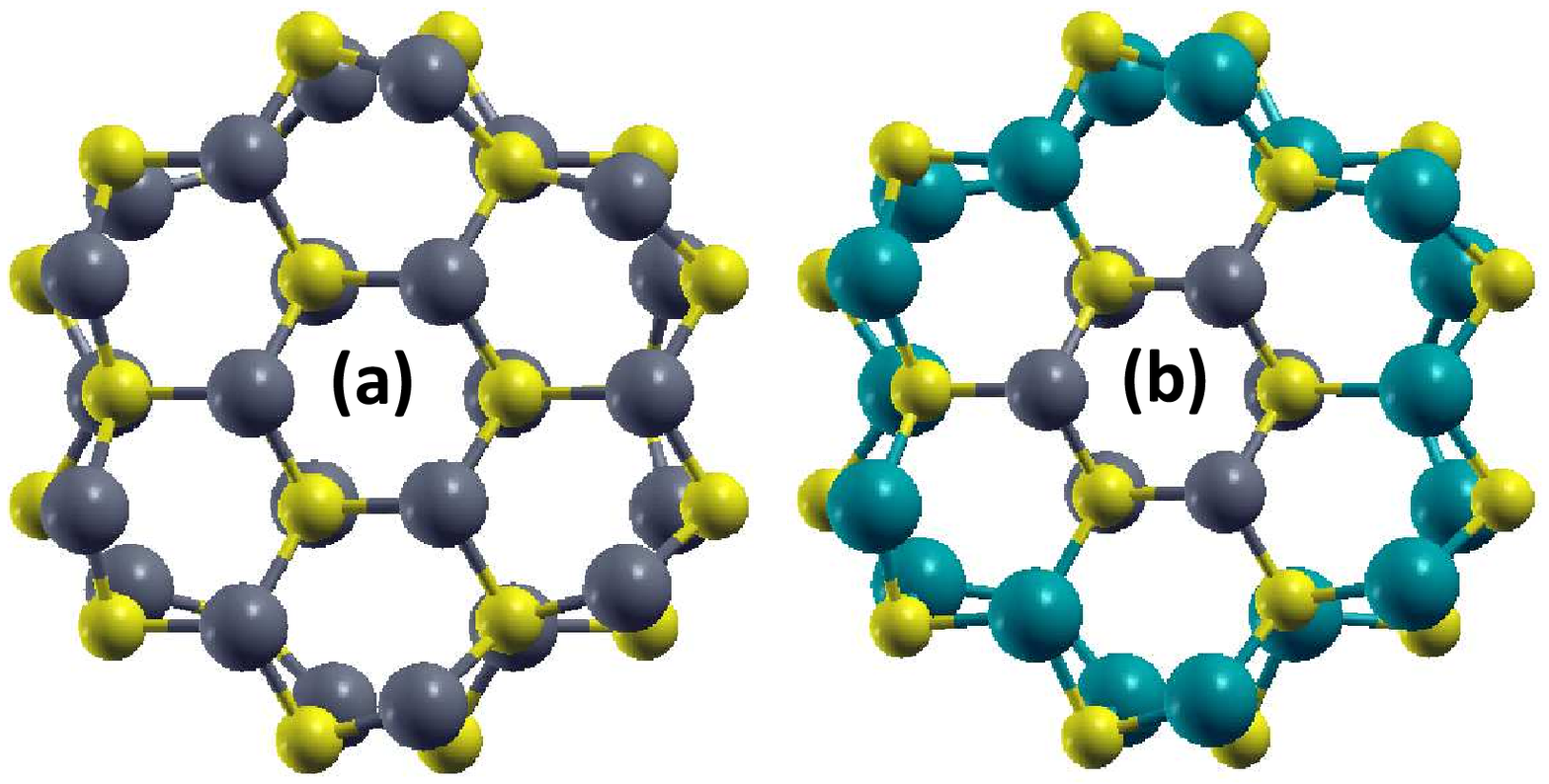}
        } 
\caption{(Color online) DFT-optimized structure of (a) ZnS and (b) ZnS/CdS nanowire. Each structure contains 48 atoms. Color codes are same as figure 1.}
\end{figure}

\begin{figure}
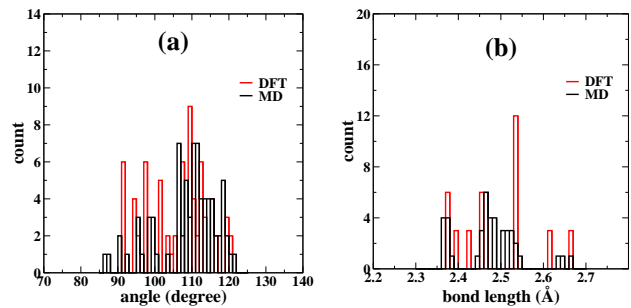

\centering
         \subfigure
         {
         \includegraphics [height=40mm]{anglecompare.eps}
         }
         \subfigure
         {
         \includegraphics [height=40mm]{bondcompare.eps}
         } 
\caption{(Color online) Comparison of (a) angle and (b) bond length distribution calculated by MD and DFT method.}
\end{figure}

\begin{figure}
\centering
        \subfigure
        {
        \includegraphics[height=65mm, width=65mm]{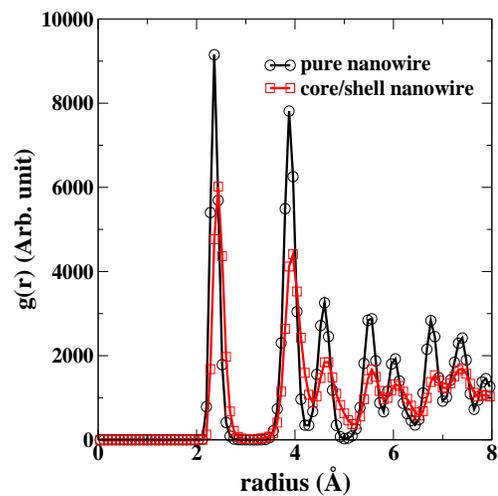}
        } 
\caption{(Color online) RDF of ZnS and ZnS/CdS nanowires. Sharp peaks in the RDF of core/shell nanowire indicates that the core/shell nanowires remain well crystalline. }
\end{figure}

To test the applicability of the ZnS and CdS potential parameters in the core/shell nanowire, we have also performed the DFT optimization of a 1.15 nm core/shell nanowire (figure 2). The DFT calculation is done within the Perdew-Burke-Ernzerhof (PBE) corrected  generalized-gradient approximation (GGA) ~\cite {perdew1996generalized} as implemented in the Quantum Espresso package ~\cite{giannozzi2009quantum}. The relaxation of the structure is done keeping the nanowire at the middle of a super cell. The dimensions of $a$ and $b$ of the super cell are kept such that the distances between the nanowire and its periodic images are approximately 10$\AA$. The dimension of the super cell along the $c$ axis is the same as that of the nanowire. We have optimized the dimension of $c$. The electron wave functions are expanded in the plane wave basis set with a cut-off energy of 30 Ry and plane waves with kinetic energy of up to 300 Ry are used for the charge density. DFT optimized structures are shown in figure 2. We have compared the bond length and angle distribution obtained from MD and DFT calculation in figure 3. We observe that the distributions are very similar in these two methods. This validates the accuracy of the classical MD potential. There is a slight deviation which may arise due to very small size of the nanowire considered in the DFT calculation.

\section {Results and discussions}
In figure 1, we present the equilibrated structures of a pure ZnS and ZnS/CdS core/shell nanowire. To scrutinize the internal structure, we have calculated the radial distribution function (RDF) of the core/shell nanowire (figure 4). We observe that the structure remains well crystalline at room temperature. However, the peak positions of the core/shell structures shift slightly towards the right compared with those of the pure ZnS structures due to larger lattice constants of CdS. Figure 5 shows the stress-strain relationship of the core/shell nanowires under uniaxial strain along the [0001] direction. To compute the uncertainty in the stress measurement, we divide the stress trajectory into five sub-trajectories. Errors are calculated from the standard deviations of the stress values of these five sub-trajectories. Since the errors are very small, the size of the error bars are very similar to the symbol size. So the error bars are not shown in the figure as they are not clearly visible. The nanowires can be deformed elastically up to a high strain of $\sim5-6\%$. From the linear region of the stress-strain curves, we have calculated the Young's moduli of the core/shell nanowires and the values are given in Table 1. We find that the Young's moduli of the core/shell nanowires are significantly lower than that of the pure ZnS nanowire. More specifically, the Young's moduli of shell1, shell2 and shell3 nanowires are 30\%, 53\% and 62\% lower than that of the pure ZnS nanowire. We find that the stress-strain response of the shell3 nanowire is low compared to that of the others at high strain. We repeated the simulation with a different initial configuration and found a similar response. To understand this low response of the shell3 nanowire, we have calculated the interfacial energy of the core/shell nanowires. The interfacial energy is calculated using following equation:

\begin{equation}
E_{interface}=E_{core/shell}^{ZnS/CdS}-E_{core}^{ZnS}-E_{shell}^{CdS}
\end{equation}

where $E_{core/shell}^{ZnS/CdS}$ is the energy of the core/shell nanowire. $E_{core}^{ZnS}$ is the energy of core region (equivalent to the core of the ZnS/CdS core/shell nanowire) taken from pure ZnS nanowire. Similarly, $E_{shell}^{CdS}$ is the energy of the shell region (equivalent to the shell of the ZnS/CdS core/shell nanowire) taken from pure CdS nanowire. Schematic diagram of the right hand side of equation 1 is shown in figure 6. So for a shell3 nanowire (total six layers): second term of equation 1 is the energy of the shell3 nanowire, third term is the energy of three inner most layers of the pure ZnS nanowire and fourth term is the energy of the three outer most layers of the pure CdS nanowire. The interfacial energies of the core/shell nanowires are 1102, 2631, 2724, 1873 and 1153 kcal/mol for shell1, shell2, shell3, shell4 and shell5 nanowire, respectively. Positive value of the interfacial energy indicates that the pure nanowires are energetically more favorable compared to the core/shell nanowires. We find that interfacial energy of the core/shell nanowires is not monotonic. Initially it increases with increasing shell material, reaches to a maxima and then finally decreases with shell material. To confirm this non-monotonic trend of the interfacial energy, we have calculated the interfacial energies of bigger core/shell nanowires (containing total number of seven atomic layers.) The interfacial energies are 1686, 3086, 2922, 3130, 1944 and 1388 kcal/mol for shell1, shell2, shell3, shell4, shell5 and shell6 nanowire, respectively. Thus, the core/shell nanowires having very small or very large shell region are energetically more favorable than the nanowires having core and shell region comparable in size. The core/shell nanowires of the last category are energetically less favorable because of their larger effective defect regions which can be understood from the figure 7. Higher interfacial energies of these core/shell nanowires easily create defects at high strain. Thus, we observe lower stress-strain response in the shell3 nanowire at high strain (figure 5). To confirm the effect of this high interfacial energy at high strain, we have calculated the stress-strain responses of bigger ZnS/CdS core/shell nanowires (containing total number of seven atomic layers). The results are shown in figure 8. Note that shell2, shell3 and shell4 nanowires show relatively low stress-strain responses at higher strain since these nanowires have higher interfacial energies as we discussed earlier. However, these nanowires are stable at low strain and show a linear stress-strain response. Note that the Young's modulus value monotonically decreases with increasing shell material and gradually converges to the Young's modulus value of the pure CdS nanowire since ZnS is much stiffer than the CdS. We conjecture that Young's modulus value also may show a similar non-monotonic behavior in a core/shell nanowire having core and shell material of same Young's modulus value.


\begin{figure}
\centering
        \subfigure
        {
        \includegraphics[height=65mm, width=65mm]{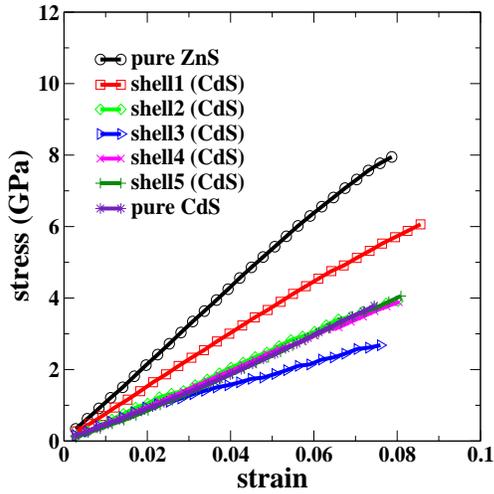}
        } 
\caption{(Color online) stress-strain response of the ZnS/CdS core/shell nanowires.}
\end{figure}

\begin{figure}
\centering
        \subfigure
        {
        \includegraphics[height=35mm,width=85mm]{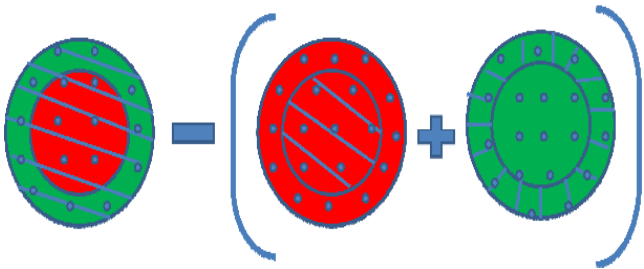}
        }  
\caption{(Color online) Schematic diagram of the method to calculate the interfacial energy of the ZnS/CdS core/shell nanowires. The energies of the region marked by blue lines are considered for the interfacial energy calculation. Red and green color represent the ZnS and CdS material, respectively. First, second and third diagram represent $E_{core/shell}^{ZnS/CdS}$, $E_{core}^{ZnS}$ and $E_{shell}^{CdS}$, respectively (see text).}
\end{figure}

\begin{figure}
\centering
        \subfigure
        {
        \includegraphics[height=70mm,width=85mm]{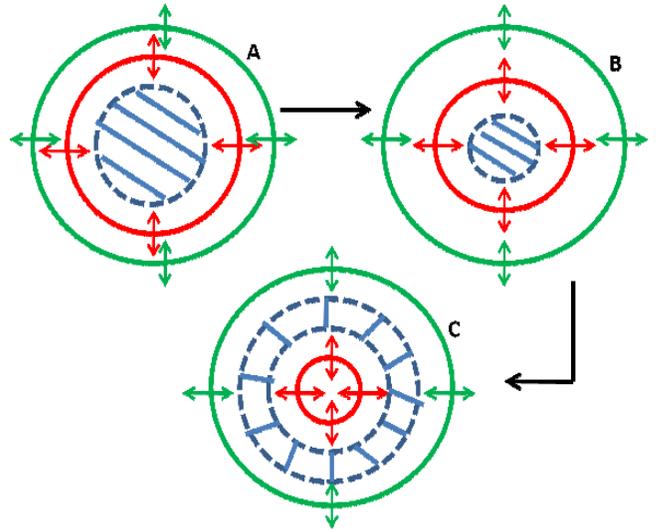}
        }  
\caption{(Color online) Schematic diagram to explain the non-monotonic behavior of the interfacial energy of the ZnS/CdS core/shell nanowires. Green and Red circles are the surface and interface, respectively. Green and red arrows represent the range of the surface and interface effects, respectively. The regions marked by the blue lines inside the dotted circles are not affected by either the surface or interface region. These regions are perfect and energetically more favorable. Black arrows indicate increment of the shell material in going from configuration A to B to C. Configuration A, B and C have very small, medium and very large shell region, respectively. Note that configuration B has the lowest perfect region which makes it energetically less favorable.}
\end{figure}

\begin{figure}[h!]
\centering       
        \includegraphics[height=65mm, width=65mm]{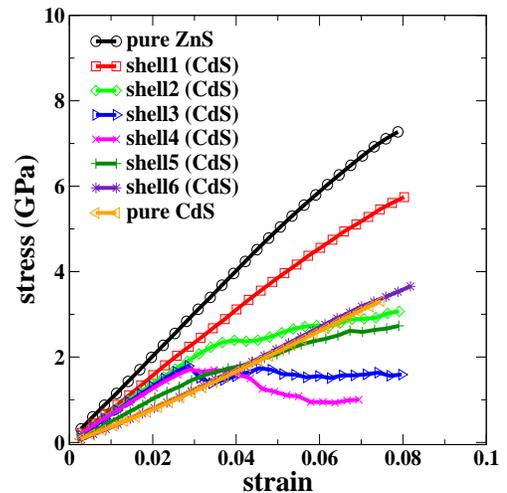}
\caption{(Color online) stress-strain responses of the ZnS/CdS core/shell nanowires containing total number of seven atomic layers. Note that shell2, sell3 and shell4 nanowires have low stress-strain responses at higher strain due to their large interfacial energies.}
\end{figure}

\begin{figure}
\centering
        \subfigure
        {
        \includegraphics[height=55mm, width=70mm]{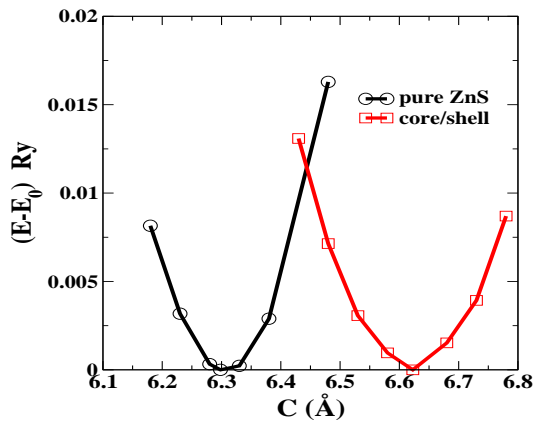}
        } 
\caption{(Color online) strain energy of ZnS and ZnS/CdS nanowires as a function of lattice constant C. Strain energy is calculated using DFT method.}
\end{figure}

To confirm the reduction of the Young's modulus in the core/shell nanowires, we have also calculated the Young's modulus of a 1.15 nm pure ZnS and ZnS/CdS core/shell nanowire (figure 2) using first principle DFT method. The nanowires are deformed along the [0001] directions. The strain energies as a function of strain for a pure and a core/shell nanowires are shown in figure 9. The Young's modulus ($Y$)is calculated by using the formula 
 
\begin{equation} Y=\frac{1}{V_0} \frac {\boldsymbol\delta^2E(\boldsymbol \epsilon)}{\boldsymbol \delta \boldsymbol \epsilon^2}\end{equation}

where $V_0, E$ and $\epsilon$ are the initial volume, strain energy and strain, respectively. The Young's modulus value of the pure ZnS nanowire is 168 GPa which is in good agreement with the previously reported value of 175 GPa~\cite{chen2009electronic} for the pristine ZnS nanowire of same size. The Young's modulus value of the ZnS/CdS core/shell nanowire (107 GPa) is much smaller than that of the pure ZnS nanowire (168 GPa). This observation validates the results obtained from the classical MD simulations.

\begin{figure}
\centering
        \subfigure
        {
        \includegraphics[height=45mm]{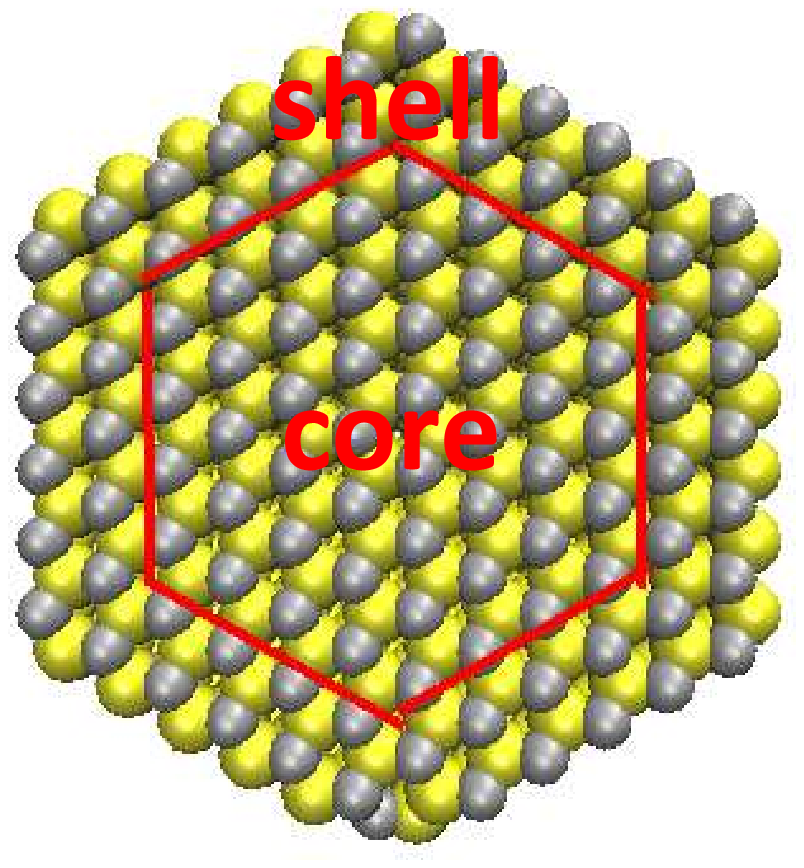}
        }         
\caption{(Color online) The core and shell region of a pure ZnS nanowire. We have calculated Young's modulus for the core and shell region separately. (see text)}
\end{figure}

\begin{figure}
\centering
        \subfigure
        {
        \includegraphics[height=65mm, width=65mm]{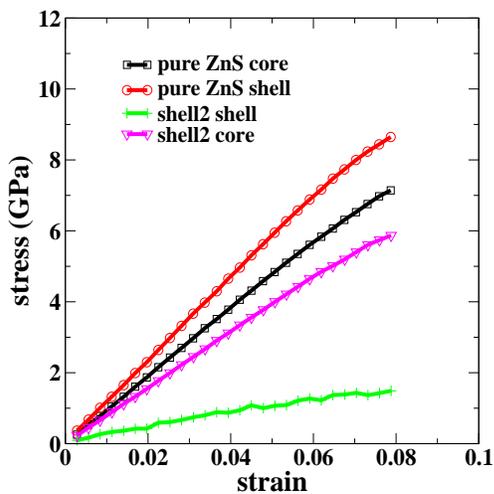}
        } 
        
\caption{(Color online) stress-strain response of the core and shell region of ZnS and ZnS/CdS nanowire. See text for the definition of shell region of ZnS nanowire.}
\end{figure}

To understand the microscopic origin behind the significant reduction in the Young's modulus value of the core/shell nanowire, we have calculated the Young's moduli values of the core and shell separately using classical MD simulation. To define a core and shell region in a pure ZnS nanowire, we decomposed the nanowire into two parts as shown in figure 10. The surface atoms have usually larger energy compared to the energy of the core atoms due to the relatively lower coordination number of the surface atoms. Thus, the surface atoms move their position to minimize the energy which is called surface relaxation. Because of the rearrangements of the surface atoms, average bond length at the surface region and hence the average bond energy of the surface atoms may increase or decrease. Thus, the Young's modulus value of the shell region is different from that of the core region of the nanowire. We measure the average bond lengths to be $2.35\pm 0.04$ $\AA$ and $2.38\pm 0.003$ $\AA$ for the surface and core regions, respectively. Note that the fluctuations in the bond length of the core region is much lower than those of the surface region which suggests that the core region remain almost unchanged. A similar kind of bond length decrease for surface atoms in ZnO nanowires was reported by Espinosa {\it et al.}~\cite{agrawal2008elasticity}. This decrease in the average bond length at the surface region increases the average bond energy of the surface atoms and hence the Young's modulus value of the surface region is also larger than that of the core region as can be seen from the stress-strain responses in figure 11. A core/shell nanowire is different from the pure nanowire due to the fact that its shell region is composed of different kinds of atoms and hence the Young's modulus value of the shell region is expected to be different from the core region. From figure 11, we observe that the slope of the stress-strain curve (Young's modulus value) of the shell region is much smaller than that of the core region of a shell2 nanowire. As a result, Young's modulus value of the entire nanowire is also smaller than that of the pure nanowire. Thus, the presence of a relatively less stiff shell at the surface region can reduce the Young's modulus value of the nanowire significantly. 

\begin{table}
\centering 
    \begin{tabular}{lc@{\hspace{0.1cm}}c@{\hspace{0.1cm}}c@{\hspace{0.1cm}}c@{\hspace{0.1cm}}c@{\hspace{0.1cm}}c@{\hspace{0.1cm}}r}
    \hline
    ZnS/CdS & ZnS & CdS & Expected & Obtained & Extra  \\ 
    nanowire        & atoms & atoms & Y (GPa) & Y (GPa) & reduction  \\         \hline
    ZnS    & 100\% & 0\% & - & 107.6 & - \\
    shell1 & 71\% & 29\% & 89.2 & 75.2 & 16\% \\ 
    shell2 & 46\% & 54\% & 73.4 & 54.8 & 25\% \\ 
    shell3 & 26\% & 74\% & 60.7 & 45.3 & 25\% \\ 
    CdS    & 0\% & 100\% & - & 44.2 & - \\
    \hline
    \end{tabular}
\caption{Expected and observed Young's modulus value of ZnS/CdS nanowires}
\end{table} 

A shell1 nanowire consists of 71\% ZnS and 29\% CdS atoms. Thus, if 71\% and 29\% contribution comes from ZnS and CdS atoms, respectively, the Young's modulus of the nanowire would be 89.6 GPa. But we observe that the Young's modulus of the nanowire has decreased further. The difference between the expected and observed Young's moduli values of various nanowires are given in Table 1. This extra reduction of the Young's modulus value suggests that the core-shell interface must be playing an important role in determining the stiffness of the nanowires. To understand the origin of this enhanced reduction of the Young's modulus, we have calculated the binding energy between the core and shell region. The binding energy is defined as 
 
\begin{equation}\boldsymbol \delta E = E_{nanowire} - E_{core} - E_{shell}\end{equation}

The binding energies between the core and shell region are -16112 kcal/mol and -9578 kcal/mol for pure ZnS and shell2 core/shell nanowire, respectively. Because of this lower binding energy, relatively less force is required to stretch the core/shell nanowire compared to the force required for the pure nanowire. As a result, the Young's modulus value of the core/shell nanowire decreases. Note that the stiffness (slope of the stress-strain response in figure 11) of the ZnS core of the ZnS/CdS core/shell nanowire has decreased further compared to that of the same ZnS core of the pure ZnS nanowire. This is because the ZnS core of a ZnS/CdS core/shell nanowire is weakly bound to the shell region but the core region of a pure ZnS nanowire is more strongly bound to the shell region.

\begin{figure}
\centering
        \subfigure
        {
        \includegraphics[height=65mm, width=65mm]{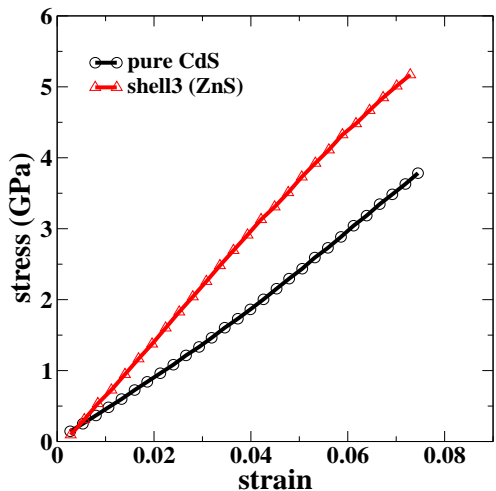}
        } 
\caption{(Color online) stress-strain response of CdS and shell3 CdS/ZnS nanowire.}
\end{figure}

To verify the effect of a stiffer shell at the surface region of a nanowire, we have also calculated the Young's modulus value of a CdS/ZnS core/shell nanowire. From figure 12, we observe that the Young's modulus value of the CdS/ZnS core/shell nanowire has increased significantly from that of the pure CdS nanowire. In this CdS/ZnS nanowire, the observed Young's modulus value (77.1 GPa) is 15\% lower than the expected value (91.1 GPa) because of the lattice mismatch at the interface of core and shell region. This observation allows us to conclude that the stiffness of a nanowire can be tuned in a wide range by depositing a suitable shell at the nanowire surface.

\begin{figure}
\centering
        \subfigure
        {
        \includegraphics[height=65mm, width=65mm]{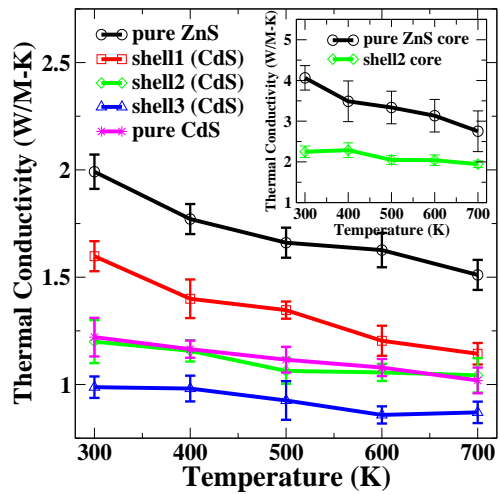}
        } 
\caption{(Color online) Thermal conductivity of ZnS, CdS and ZnS/CdS nanowire as a function of temperature. Inset shows the thermal conductivity of the core regions of pure ZnS and ZnS/CdS core/shell nanowires.}
\end{figure}

Thermal conductivity is an important parameter in designing efficient nanodevices. To investigate the influence of a thin shell on the thermal conduction of nanowires, we have calculated the thermal conductivities of the pure ZnS and ZnS/CdS core/shell nanowires. The thermal conductivity of the nanowire is calculated by an approach originally developed by Muller-Plathe~\cite{muller1997simple} which is based on the reverse non-equilibrium molecular dynamics (RNEMD) method. The nanowire is divided into few slabs along the growth direction (Z direction). One slab at the middle part of the nanowire is considered to be the heat source and two slabs, one at each end are the sink regions. The energy is transferred from the cold region to the hot region by exchanging the velocities between the hottest atom of the cold region and the coldest atom of the hot region at certain intervals. At steady state, the average energy flux is evaluated as 

 \begin {equation}j_{z}= \frac{1}{2tA}\sum_{transfers}\frac{m}{2}(v^2_{hot}-v^2_{cold})\end {equation}

Where $A$, $t$ and $transfers$ are the area of the nanowire, duration of the simulation and number of transfers during the simulation time, respectively. $v_{hot}$ and $v_{cold}$ are the velocities of the hottest and coldest atom of equal mass $m$. This constant heat flux creates a temperature gradient in the nanowire. Thermal conductivity is measured by taking the ratio of imposed thermal flux and the temperature gradient at the steady state. 

\begin {equation}k=-\frac{j_z}{<\frac{dT}{dZ}>} \end{equation}

This method has been used extensively to compute the thermal conductivity of different systems ~\cite{alaghemandi2009thermal, lin2013effect, osman2001temperature} \\

In figure 13, we have shown the thermal conductivities of pure ZnS and ZnS/CdS core/shell nanowires as a function of temperature. To compute the uncertainty in the thermal conductivity measurement, the temperature gradient trajectory is divided into five sub-trajectories. We get thermal conductivity for each sub-trajectory by calculating temperature gradient and thermal flux for each sub-trajectory. Then, we compute standard deviations of these thermal conductivities to get the error bars. Note that the thermal conductivities shown in figure 13 are the thermal conductivities of the core/shell nanowires with particular lengths we have used. Since the lengths of the nanowires are much smaller than the phonon mean free path, finite size effects would be there even though we have used periodic boundary condition. Muller-Plathe {\it et al.}~\cite{alaghemandi2009thermal} observed a strong length dependence of the thermal conductivities in carbon nanotubes even while using periodic boundary condition. Thus, thermal conductivity values reported here are smaller than the thermal conductivities of very long core/shell nanowires. As expected, the thermal conductivity decreases with temperature because of lower phonon mean free path at higher temperature. Note that the thermal conductivity of the ZnS/CdS core/shell nanowire is significantly lower than that of the pure ZnS nanowire at each temperature. More precisely, the thermal conductivities of the shell1 nanowire is 22\% lower than that of the pure ZnS nanowire at 300 K. For further microscopic insight, we have computed the thermal conductivities of the core regions of pure ZnS and shell2 nanowires (inset of figure 13). Note that the thermal conductivity of the core region of the pure ZnS nanowire is larger than that of the entire nanowire. This is because of no surface scattering in the core region since thermal conductivity has been measured in the presence of the shell region. Interestingly, thermal conductivity of the same ZnS core region is decreased significantly when the ZnS shell is replaced by CdS shell (in a shell2 nanowire). This happens due to the lattice mismatch at the ZnS-CdS interface region. Thus, lattice mismatch at the core-shell interface plays an important role in determining the thermal conductivity of the core/shell nanowires. The effect of the interface region is clearly observed in a shell3 nanowire where the thermal conductivity of the nanowire is lower than the thermal conductivity of each of its constituent materials. 

\begin{figure}
\centering
        \subfigure
        {
        \includegraphics[height=65mm, width=65mm]{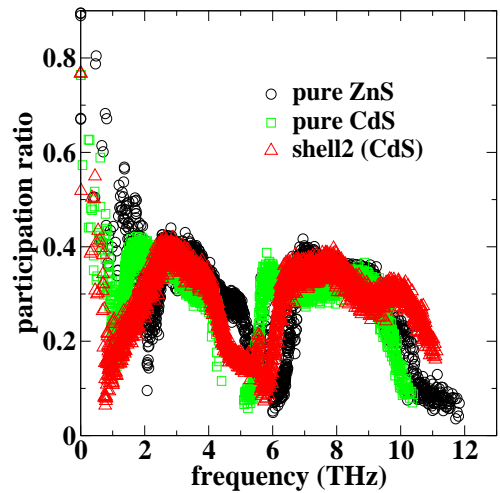}
        } 
\caption{(Color online) Participation ratio of the phonon modes of the nanowires.}
\end{figure}

To understand this significant reduction in the thermal conductivities of the core/shell nanowires, we have calculated the participation ratio of the phonon modes of the nanowires. Participation ratio $P_\lambda$ of an eigen-mode $\lambda$ is defined as

\begin{equation} P_\lambda ^{-1}=N\sum_i(\sum_\alpha \epsilon_{i\alpha,\lambda}^* \epsilon_{i\alpha,\lambda})^2\end{equation}    

where $\epsilon_{i\alpha,\lambda}$ is the eigen vector of the $\lambda^{th}$ mode. $i$ sums over the number of atoms, N. $\alpha$ sums over $x$, $y$, and $z$. The value of the participation ratio equals to $1$ and $1/N$ for completely delocalized mode and completely localized mode, respectively. The participation ratio of the pure ZnS, CdS and shell2 ZnS/CdS nanowire are shown in figure 14. We chose a small segment of the nanowire for phonon frequency calculations. The segment contains six layers of atoms (total 1296 atoms) along the longitudinal direction and the cross-sectional area is same as that of the original nanowire. Vibrational mode analysis is done using the fix-phonon~\cite{kong2011phonon} code implemented in LAMMPS~\cite{lammps}. From figure 14, we observe that the participation ratio of the shell2 phonon modes are significantly lower than that of the pure ZnS and CdS phonon modes, in particular at the low frequency (below 2 THz) range.  
Thus, the low frequency phonon modes in the core/shell nanowires are more localized. Localized phonon modes are less efficient in thermal energy conduction compared to the delocalized phonon modes~\cite{bodapati2006vibrations, chen2010remarkable}. But the low frequency phonon modes mainly contribute to the heat conduction in semiconductors. Thus, the localization of the low frequency phonon modes causes significant reduction in the thermal conductivities of the ZnS/CdS core/shell nanowires.

\begin{figure}
\centering
        \subfigure
        {
        \includegraphics[height=65mm,width=65mm]{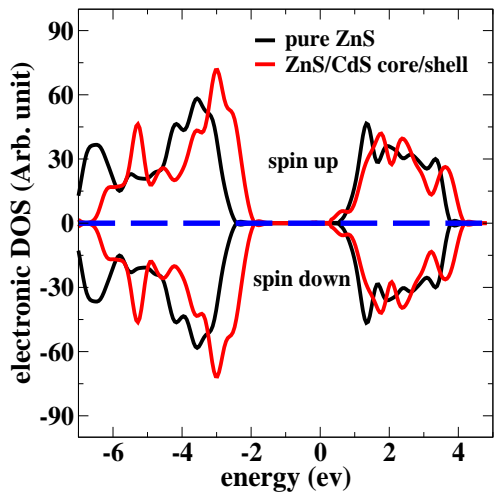}
        }  
\caption{(Color online) Electronic densities of states of pure ZnS and ZnS/CdS core/shell nanowire.}
\end{figure}

Lower thermal conductivities of the ZnS/CdS core/shell nanowires suggest that they could be better candidates for thermoelectric applications. The figure of merit (Z) of a thermoelectric material can be improved by increasing electrical conductivity and decreasing thermal conductivity of the material. But the electronic contribution to the thermal conductivity increases with the electrical conductivity of the material. So the electrical and thermal conductivity compete with each other in determining the figure of merit (Z) of a thermoelectric material. Phonons have more contribution than the electrons in the thermal conductivities of semiconductors. Thus, one way to optimize the thermoelectric efficiency of a semiconducting material is to minimize the lattice thermal conductivity and increase the electrical conductivity. We have already observed that the lattice thermal conductivity is reduced in the ZnS/CdS core/shell nanowires due to the localization of long-wavelength phonon modes. To investigate the electrical conductivity of the ZnS/CdS core/shell nanowires, we have  calculated the electronic densities of states of the nanowires (figure 15) corresponding to the structures shown in figure 2. We find the band gap to be 2.93 eV and 2.15 eV for pure ZnS and ZnS/CdS core/shell nanowire, respectively. This calculated band gap of the pure ZnS is in good agreement with the previously reported values of 2.9 eV~\cite{chen2009stability}, 2.8 eV~\cite{pan2008semiconductor} for pristine ZnS nanowire of same size. Note that the calculated band gap of the pure ZnS nanowire is lower than the experimental band gap (3.77 eV~\cite{ong2001optical}) of bulk wurtzite ZnS. However, the band gap of bulk wurtzite ZnS predicted by DFT method is 2.13 eV ~\cite{chen2009stability}, 2.14 eV~\cite{pan2008semiconductor}, 1.93 eV~\cite{zhang2008first} which are much lower than the experimental value. The underestimation of the band gap value is because of the well known limitation of the DFT method~\cite{godby1988self}. So our calculated band gap value of ZnS nanowire is larger than the bulk ZnS band gap predicted by the DFT method. This is consistent with the quantum confinement effect. We find that the band gap of ZnS/CdS core/shell nanowire decreases by 0.78 eV compared to the band gap of the pure ZnS nanowire which increases the electrical conductivity of the ZnS/CdS core/shell nanowires. Note that the calculated band gap reduction may not be very accurate due to the limitation of DFT calculation. A more accurate method such as GW approximation is required for predicting the band gap reduction correctly~\cite{hybertsen1986electron}. Thus, lower lattice thermal conductivity and higher electrical conductivity make the ZnS/CdS core/shell nanowires  better suited for thermoelectric application compared to the pure ZnS nanowires.

\begin{figure}
\centering
       \centering
        \subfigure
        {
        \includegraphics[height=70mm, width=63mm]{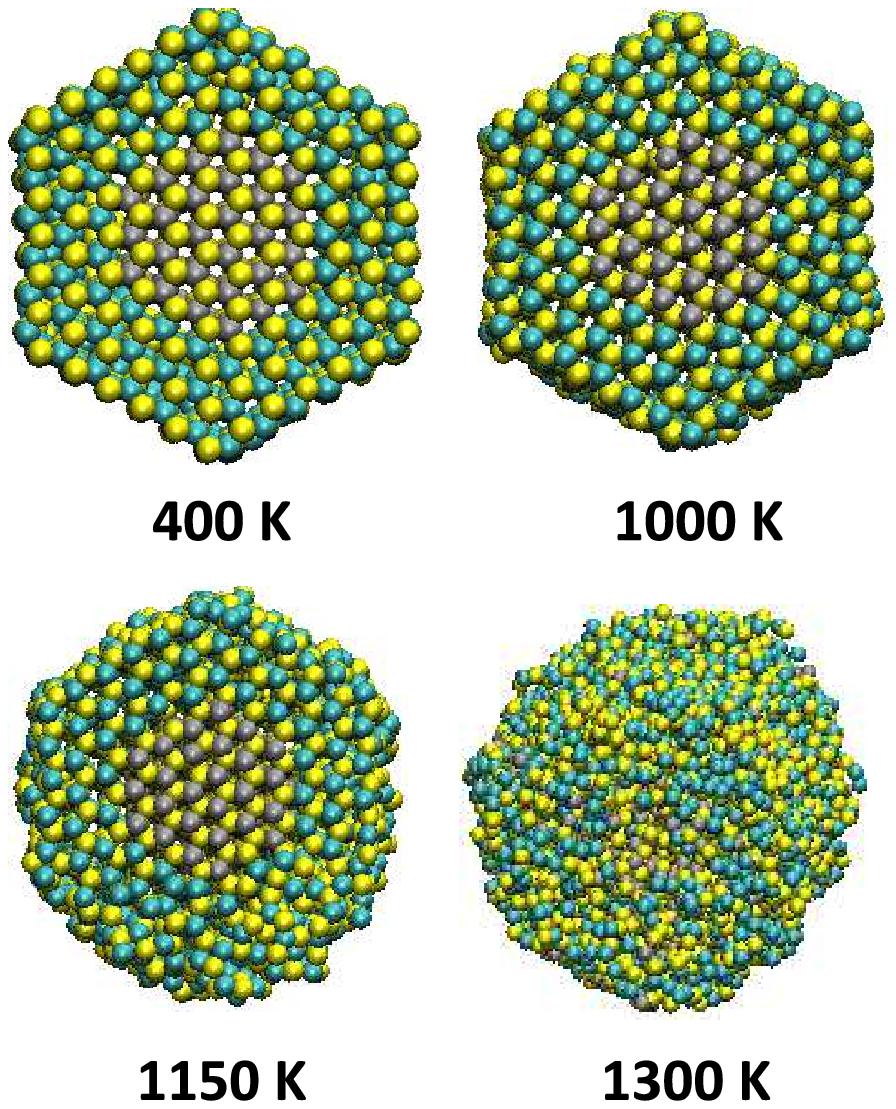}
        } 
\caption{(Color online)  Cross-sectional views of the shell3 ZnS/CdS nanowire at different temperatures. Note that the shell region melts at 1150 K temperature. However, core region preserves its geometry at this temperature. Core region melts at 1300 K. Color codes are same as figure 1.}
\end{figure}

\begin{figure}
       \centering
        \subfigure
        {
        \includegraphics[height=70mm, width=63mm]{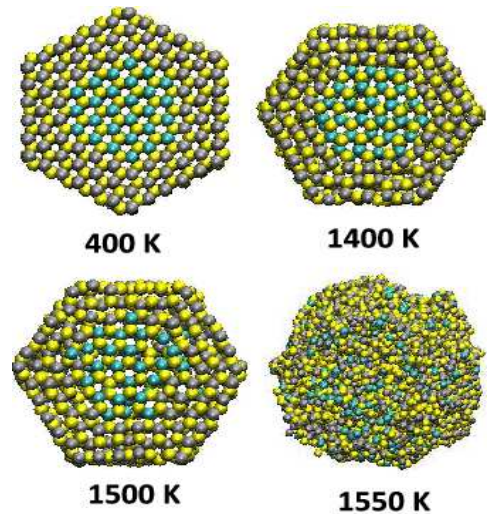}
        } 
\caption{(Color online)  Cross-sectional views of the shell3 CdS/ZnS nanowire at different temperature. In this case, the interface region starts to melt at 1500 K temperature. However, the surface region preserves its hexagonal geometry at this temperature. Color codes are same as figure 1.}
\end{figure}

Lower value of thermal conductivity makes the core/shell nanowire more efficient for building devices where large heat loss is not desired.
However, coating of a shell on top of the nanowire may affect the melting temperature of the core/shell nanowire, another important parameter which needs attention. Recently Sun {\it et al.} investigated the melting behavior of metallic core/shell and alloyed nanoparticles extensively~\cite {huang2014tunable, huang2012pt, huang2013insight, huang2013thermal, huang2012two}. They found that the melting behavior of core/shell nanoparticles significantly differs from that of its constituent materials. Thus, to investigate the effect of shell region on the melting behavior of core/shell nanowires, we gradually heat up the nanowires from 300 K to 2200 K with a temperature increment of 50 K. At each temperature, the system is equilibrated for 500 ps. Snapshots of a shell3 ZnS/CdS core/shell nanowire at different temperatures are presented in figure 16. We find that the melting initiates at the surface of the core/shell nanowire. The shell region starts to melt at a lower temperature (1150 K), at which the core region remains perfectly crystalline as can be seen in figure 16. The structure becomes completely disordered at 1300 K. However, a pure ZnS nanowire with the same diameter remains perfectly crystalline at 1300 K (snapshots not shown) which suggests that the melting temperature of a ZnS/CdS core/shell nanowire is lower than that of the pure ZnS nanowire. This can be understood from the following fact: The shell region begins to melt first because CdS has lower melting temperature compared to the melting temperature of ZnS. Once the shell is melted, the crystalline ZnS core also starts to melt at a temperature which is lower than the melting temperature of a pure ZnS nanowire since the size of the core region is smaller than that of a pure ZnS nanowire. Thus, coating of a nanostructure by a material of relatively lower melting temperature leads to a reduction in the melting temperature of the core/shell nanostructure. Here we should mention that an accurate estimation of the melting temperature of a nanostructure is difficult in the MD approach because of the following reason: To calculate the melting temperature correctly, one should separately calculate the Gibb's free energy for solid and liquid phases as a function of temperature. The intersection of the plots of the two free energies would give the melting temperature. But this method can not be applied to nanoscale systems due to the difficulty in defining the liquid phase of a nanoscale system. However, our results suggest that the melting temperature of a ZnS/CdS core/shell nanowire should be lower than that of a pure ZnS nanowire with the same diameter. We have also investigated the melting behavior of CdS/ZnS core/shell nanowires, where the shell material has higher melting temperature than the core material. We find that melting of CdS/ZnS nanowire initiates at the core-shell interface region (figure 17) unlike the ZnS/CdS nanowires where melting initiates at the surface region. This is so because ZnS has much higher melting temperature than  CdS. Note that the CdS core preserves its crystalline structure at 1400 K (figure 17) where a pure CdS nanowire melts (snapshots not shown) suggesting that the nanostructures can be protected at a temperature higher than its melting temperature by appropriate coating. Thus, our results suggest that along with the stiffness and thermal conductivity, melting temperature of nanostructure also can be tuned by appropriate choice of the coating material.

\section{Conclusions}

In summary, we have employed fully atomistic MD simulation to investigate the mechanical and thermal properties of ZnS/CdS core/shell nanowires. We observe that coating of a few atomic layers of CdS on top of the ZnS nanowires leads to a significant reduction in the Young's moduli of the nanowires. We attribute this strong decrease in Young's modulus of nanowires to the lattice mismatch at core-shell interface. We have calculated the interfacial energies of the core/shell nanowires. Non-monotonic behavior of the interfacial energy suggests that core/shell nanowires with very small or very high shell region are energetically more favorable. The interfacial energy determines the stability of the core/shell nanowires at high strain. We have also shown that the Young's modulus of a nanowire can be increased by depositing a relatively stiffer shell on the nanowire surface. Thermal conductivity of ZnS/CdS core/shell nanowires is also lower than that of the pure ZnS nanowires as the low frequency phonon modes are more localised in the core/shell nanowires. We find that the band gap of the ZnS/CdS core/shell nanowire decreases significantly from that of the pure ZnS nanowire. Thus, lower thermal conductivity and higher electrical conductivity make the ZnS/CdS nanowires better suited for thermoelectric devices compared to the pure ZnS nanowires. We have also shown that the melting temperature of the core/shell nanowires can be tuned in a wide range by an appropriate choice of the coating material. Knowledge of the mechanical and thermal behavior of the core/shell nanostructures will help in designing efficient nanodevices. This study will motivate the experimentalists to carry out further investigations on mechanical and thermal properties of semiconducting core/shell nanostructures. 

\section{Acknowledgments}  

We thank DST, India for financial support. T.M. thanks Council of Scientific and Industrial Research (CSIR), India for fellowship.


%

\end{document}